# Network Deployment for Maximal Energy Efficiency in Uplink with Zero-Forcing

Andrea Pizzo*, Daniel Verenzuela†, Luca Sanguinetti*‡, Emil Björnson†
*Dipartimento di Ingegneria dell'Informazione, University of Pisa, Pisa, Italy
†Department of Electrical Engineering (ISY), Linköping University, Linköping, Sweden.
‡Large Networks and System Group (LANEAS), CentraleSupélec, Université Paris-Saclay, Gif-sur-Yvette, France

*Abstract*—This work aims to design a cellular network for maximal energy efficiency (EE). In particular, we consider the uplink with multi-antenna base stations and assume that zero-forcing (ZF) combining is used for data detection with imperfect channel state information. Using stochastic geometry and a new lower bound on the average per-user spectral efficiency of the network, we optimize the pilot reuse factor, number of antennas and users per base station. Closed-form expressions are computed from which valuable insights into the interplay between the optimization variables, hardware characteristics, and propagation environment are obtained. Numerical results are used to validate the analysis and make comparisons with a network using maximum ratio (MR) combining. The results show that a Massive MIMO setup arises as the EE-optimal network configuration. In addition, ZF provides higher EE than MR while allowing a smaller pilot reuse factor and a more dense network deployment.

## I. INTRODUCTION

Keeping up with the ever-growing demands for higher data throughput is the major ambition of 5G cellular networks [1]. An important question is how to evolve communication technologies to deliver higher throughputs without increasing the power consumption prohibitively [2]. This calls for new design mechanisms that allow to provide the user equipments (UEs) with high spectral efficiency at moderately low energy costs. There is a broad consensus that this can only be achieved with a substantial network densification [3]. The different approaches for this are: small-cell networks [4], [5], Massive multiple-input-multiple-output (MIMO) systems [6]–[8] and a combination thereof [9]. The former needs many more base stations (BSs) while the latter increases the hardware cost per BS. Small-cell networks guarantee lower propagation losses through base stations (BS) densification while Massive MIMO makes use of a large number $M$ of antennas per BS to serve a relatively large number $K$ of UEs with $M \gg K$. Notice that both solutions tend to increase the energy needed to run the network; that is, the energy consumed per BS times the number of BSs deployed. Recently, [9] considered the uplink (UL) of a network using MR combining for data detection under the assumption that BS locations are drawn according to a homogeneous Poisson point process (PPP). This allows a tractable analysis of network performance, from which it turns out that densification provides higher EE up to a certain point where further benefits can only be achieved through a Massive MIMO configuration.

The aim of this work is to extend the analysis in [9] to a network in which ZF combining is used rather than MR, since ZF generally provides better rates. We obtain a lower bound on the achievable EE and maximize it analytically with respect to $M$, $K$ and the pilot reuse factor. The resulting expressions reveal the fundamental interplay of these three design parameters, which are also illustrated numerically. In particular, it turns out that the use of ZF allows a higher densification of the network while using a smaller pilot reuse factor and achieving a higher EE than with MR combining. Both schemes employ almost the same optimal number of antennas per BS to approximately serve the same number of UEs, with a ratio $M/K$ at the order of 10. However, ZF is characterized by a smoother EE function, which is more robust to system changes and thus makes it a better choice.

## II. NETWORK MODEL

We consider the UL of a cellular network wherein the BSs are distributed spatially in $\mathbb{R}^2$ according to a homogeneous PPP $\Phi_\lambda$ of intensity $\lambda$ [BS/km$^2$]. Each BS has $M$ antennas and serves simultaneously $K$ single-antenna UEs over a bandwidth of $B_{\mathrm{w}}$ [MHz]. These UEs are selected at random from a (potentially) very large set of UEs in the cell. We assume that each UE connects to its closest BS such that the coverage area of a BS is its Poisson-Voronoi cell. The UEs are assumed to be uniformly distributed in the Poisson-Voronoi cell of their serving BS. Without loss of generality, we assume that the "typical UE", which is statistically representative for any other UE in the network [10], has an arbitrary index $k$ and it is connected to an arbitrary BS $j$. This means that the distance between UE $k$ in cell $j$ to its serving BS $j$, denoted by $d_{jk}^j$, is distributed as $d_{jk}^j \sim \mathrm{Rayleigh}\big(\sqrt{1/(2\pi\lambda)}\big)$ [10].

The channel response $\mathbf{h}_{lk}^j \in \mathbb{C}^M$ between UE $k$ in cell $l$ and the BS in cell $j$ is modeled as uncorrelated Rayleigh fading such that $\mathbf{h}_{lk}^j \sim \mathcal{CN}(\mathbf{0}, \beta_{lk}^j \mathbf{I}_M)$. The large-scale fading coefficient $\beta_{lk}^j$ is modeled as $\beta_{lk}^j = \Upsilon(d_{lk}^j)^{-\alpha}$, where $d_{lk}^j$ [km] is the distance between UE $k$ in cell $l$ and the BS in cell $j$, $\alpha$ is the pathloss exponent and $\Upsilon$ accounts for the pathloss at a reference distance of 1 km. The UEs use a statistical power-control policy in the UL such that the transmit power of UE $k$ in cell $j$ is $p_{jk} = P_0/\beta_{jk}^j$ with $P_0$ being a design parameter. This provides uniform average SNR to all the UEs without requiring any instantaneous channel state information.

This research has been supported by the ERC Starting Grant 305123 MORE and the Swedish Foundation for Strategic Research (SFF).

We assume that a pilot book $\mathbf{\Phi} \in \mathbb{C}^{\tau_p \times \tau_p}$ of $\tau_p \geq K$ mutually orthogonal UL sequences is used in each cell for channel estimation at the BS. To avoid pilot coordination, we assume that in each coherence block each BS $j$ picks a subset of $K$ different pilot sequences $\{\boldsymbol{\phi}_{jk}; k = 1, \ldots, K\}$ at random. This means that on average $\zeta = \tau_p/K$ of the cells reuse any given pilot sequence. This is modeled through a binary stochastic variable $a_l^j \in \{0, 1\}$, where $a_l^j = 1$ means that UE $k$ in cell $l$ uses the same pilot as the typical UE with probability $1/\zeta$. Similarly, $a_l^j = 0$ stands for no pilot contamination from cell $l$ occuring with probability $1 - 1/\zeta$.

Under the above setting, by using the assumed power-control policy, the MMSE estimator of $\mathbf{h}_{jk}^j$ at BS $j$ is [7], [9]

$$\hat{\mathbf{h}}_{jk}^j = \frac{\sqrt{\beta_{jk}^j/P_0}}{1 + \sum_{l \in \Psi_\lambda} a_l^j \left(\frac{d_{lk}^l}{d_{lk}^j}\right)^\alpha + \frac{1}{\mathsf{SNR}_p}} \frac{1}{\tau_p} \mathbf{Y}_j^p \boldsymbol{\phi}_{jk}^\star \quad (1)$$

with the received processed pilot signal given by

$$\mathbf{Y}_j^p \boldsymbol{\phi}_{jk}^\star = \underbrace{\sqrt{p_{jk}} \tau_p \mathbf{h}_{jk}^j}_{\text{Desired pilot}} + \underbrace{\sum_{l \in \Psi_\lambda} a_l^j \sqrt{p_{lk}} \tau_p \mathbf{h}_{lk}^j}_{\text{Interfering pilots}} + \underbrace{\mathbf{N}_j^p \boldsymbol{\phi}_{jk}^\star}_{\text{Noise}} \quad (2)$$

where $\mathsf{SNR}_p = \frac{P_p}{\sigma^2}$ accounts for the effective SNR during pilot signaling from UE $k$ in cell $j$ to its serving BS $j$ with $P_p$ being the effective signaling power including the pilot length $\tau_p$. The noise is distributed as $\mathbf{N}_j^p \boldsymbol{\phi}_{jk}^\star \sim \mathcal{N}_\mathbb{C}(\mathbf{0}, \sigma^2 \tau_p \mathbf{I}_M)$. The MMSE estimate $\hat{\mathbf{h}}_{jk}^j$ and the estimation error $\tilde{\mathbf{h}}_{jk}^j$ are independent random vectors, distributed as $\hat{\mathbf{h}}_{jk}^j \sim \mathcal{N}_\mathbb{C}(\mathbf{0}, \gamma_{jk}^j \mathbf{I}_M)$ and $\tilde{\mathbf{h}}_{jk}^j \sim \mathcal{N}_\mathbb{C}(\mathbf{0}, (\beta_{jk}^j - \gamma_{jk}^j)\mathbf{I}_M)$ with

$$\gamma_{jk}^j = \frac{\beta_{jk}^j}{1 + \sum_{l \in \Psi_\lambda} a_l^j \left(\frac{d_{lk}^l}{d_{lk}^j}\right)^\alpha + \frac{1}{\mathsf{SNR}_p}}. \quad (3)$$

The channel vector $\hat{\mathbf{h}}_{li}^j$ to UE $i$ in cell $l$ that uses the same pilot sequence (i.e., $a_l^j = 1$) is $\hat{\mathbf{h}}_{li}^j = \beta_{li}^j (\sqrt{\beta_{li}^l \beta_{jk}^j})^{-1} \hat{\mathbf{h}}_{jk}^j$. The above expression describes one of the key characteristics of the pilot contamination phenomenon in presence of an uncorrelated fading, namely, the two channel estimates $\hat{\mathbf{h}}_{li}^j$ and $\hat{\mathbf{h}}_{jk}^j$ are parallel vectors that only differ in scaling.

During UL payload transmission, each BS $j$ detects the $K$ desired signals of powers $\{p_{jk}\}$ by using a receive combining matrix $\mathbf{V}_j$. In this work, we consider a ZF combining such that $\mathbf{V}_j^{\mathsf{ZF}} = \hat{\mathbf{H}}_j^j((\hat{\mathbf{H}}_j^j)^{\mathsf{H}} \hat{\mathbf{H}}_j^j)^{-1}$ where $\hat{\mathbf{H}}_j^j = [\hat{\mathbf{h}}_{j1}^j \ldots \hat{\mathbf{h}}_{jK}^j]$ denotes the $M \times K$ matrix with $M > K$, which collects the estimates from all the UEs in cell $j$ to BS $j$.

III. PROBLEM STATEMENT

Let us introduce the tuple of parameters $\boldsymbol{\theta} = (\zeta, K, M)$ defined over a set $\Theta = \{\boldsymbol{\theta} : 1 \leq \zeta \leq \tau_c/K, (M, K) \in \mathbb{Z}_{++}\}$ where $\tau_c$ denotes the size of the channel coherence block such that $K\zeta \leq \tau_c$ represents the upper limit on the pilot signaling overhead. This work aims to find $\boldsymbol{\theta}$ that solves the following EE maximization problem:

$$\begin{aligned}\underset{\boldsymbol{\theta} \in \Theta}{\text{maximize}} \quad & \mathsf{EE}(\boldsymbol{\theta}) = B_{\mathsf{w}} \frac{\mathsf{ASE}(\boldsymbol{\theta})}{\mathsf{APC}(\boldsymbol{\theta})} \\ \text{subject to} \quad & \underline{\mathsf{SINR}} \geq \gamma\end{aligned} \quad (5)$$

where ASE [bit/s/Hz/km$^2$] and APC [W/km$^2$] stand for the area spectral efficiency and the area power consumption, respectively. The parameter $\gamma > 0$ in (5) is used to impose an average SE constraint of $\log_2(1+\gamma)$ in bit/s/Hz/UE, where the average is computed with respect to both BS and UE locations.

A. Area Spectral Efficiency

The ergodic UL capacity for a network (such as the one under investigation) with imperfect CSI and inter-cell interference modeled as a shot-noise process is not known yet [11]. To overcome this issue, we compute a lower bound on the average ergodic capacity of the typical UE. For notational convenience, the following quantities are introduced for $\kappa = 1, 2$:

$$\vartheta_{ji}^{(\kappa)} = \sum_{l \in \Psi_\lambda} \left(\frac{d_{li}^l}{d_{li}^j}\right)^{\kappa\alpha}. \quad (6)$$

**Theorem 1.** *When ZF combining is used and the UL powers $\{p_{jk}\}$ are chosen as $p_{jk} = P_0/\beta_{jk}^j$, the average ergodic capacity of the typical UE $k$ in cell $j$ is lower bounded by*

$$\left(1 - \frac{K\zeta}{\tau_c}\right) \mathbb{E}_d\{\log_2(1 + \underline{\mathsf{SINR}}_{jk})\} \quad (7)$$

*with* $\underline{\mathsf{SINR}}_{jk}$ *given by* (4) *at the top of the next page and the expectation* $\mathbb{E}_d\{\cdot\}$ *computes the average over UEs' positions.*

*Proof.* The proof is sketched (due to space limitations) in Appendix A. □

An average UL SE per UE in (7) can be obtained by taking the expectation with respect to the PPP $\Phi_\lambda$ and UE positions, which would require Monte Carlo simulations. The following lemma provides a tractable lower bound.

**Lemma 1.** *When ZF combining is used with $M > K$ and the UL powers $\{p_{jk}\}$ are chosen as $p_{jk} = P_0/\beta_{jk}^j$, a lower bound on the UL average ergodic SE per UE is* $\underline{\mathsf{SE}} = (1 - K\zeta/\tau_c) \log_2(1 + \underline{\mathsf{SINR}})$ *where*

$$\underline{\mathsf{SINR}} = \frac{M-K}{\underbrace{\mathsf{IN}}_{\text{Interference plus noise}} + \underbrace{\mathsf{PC}}_{\text{Pilot contamination}}} \quad (8)$$

*with* $\mathsf{PC} = \frac{M-K}{\zeta(\alpha-1)}$ *and*

$$\begin{aligned}\mathsf{IN} = & \left(K + \frac{1}{\mathsf{SNR}}\right)\left(1 + \frac{2}{\zeta(\alpha-2)} + \frac{1}{\mathsf{SNR}_p}\right) \\ & + \frac{2K}{\alpha-2}\left(1 + \frac{1}{\mathsf{SNR}_p}\right) + \frac{K}{\zeta}\left(\frac{4}{(\alpha-2)^2} + \frac{1}{\alpha-1}\right) \\ & - K\left(1 + \frac{1}{\zeta(\alpha-1)}\right).\end{aligned} \quad (9)$$

*Proof.* The proof is sketched (due to space limitations) in Appendix B. □

$$\underline{\mathsf{SINR}}_{jk} = \frac{M-K}{\left(K + \frac{1}{\mathsf{SNR}} + \sum_{i=1}^{K}\vartheta_{ji}^{(1)}\right)\left(1 + \frac{1}{\mathsf{SNR}_p} + \frac{1}{\zeta}\vartheta_{jk}^{(1)}\right) + \frac{M-K}{\zeta}\vartheta_{jk}^{(2)} - \left(K + \frac{1}{\zeta}\sum_{i=1}^{K}\vartheta_{ji}^{(2)}\right)} \quad (4)$$

By using Theorem 1, the ASE is obtained as

$$\mathsf{ASE} = \lambda K \, \underline{\mathsf{SE}}. \quad (10)$$

The tightness of the lower bound in (10) will be validated later on by means of Monte Carlo simulations. Observe that the numerator of $\underline{\mathsf{SINR}}$ in (8) scales with $M-K$ since each BS sacrifices $K$ degrees of freedom for interference suppression within the cell. The term PC scales also with $M-K$ and accounts for pilot contamination due to UEs that use the same pilot sequence as the typical UE. Many of the interference terms in IN increase with $K$ since having more UEs lead to more interference per cell. Comparing the above expressions with those obtained in [9] with MR combining, we notice that the numerator scales with $M$ rather than $M-K$ since MR combining suppresses interference and noise by amplifying the signal of interest using the full array gain of $M$. The interference from other cells is the same for both schemes except for the extra negative term in (9), which makes ZF combining preferable to MR combining whenever the reduced interference is more substantial than the loss in array gain. The pilot contamination term scales also with $M$ (as the useful signal) rather than $M-K$ as in PC since MR combining does not benefit from interference suppression.

### B. Area Power Consumption

The APC of the network is computed as in [9] using the network parameters given in Table I and taking into account that ZF combining is used for detection. This leads to:

$$\mathsf{APC} = \underbrace{\lambda \overline{\mathsf{APC}}}_{\substack{\text{Power consumed} \\ \text{by digital processing}}} + \underbrace{B_\mathrm{w} \mathcal{A} \mathsf{ASE}}_{\substack{\text{Power consumed} \\ \text{by coding, decoding and backhaul}}} \quad (11)$$

where $\overline{\mathsf{APC}}$ has a polynomial structure in $\zeta$, $M$ and $K$:

$$\overline{\mathsf{APC}} = \mathcal{C}_0 + \mathcal{C}_1 K - \mathcal{C}_2 K^2 \zeta + \mathcal{C}_3 K^3 + \mathcal{D}_0 M + \mathcal{D}_1 MK + \mathcal{D}_2 MK^2 \quad (12)$$

where we have defined $\mathcal{C}_1 = \bar{\mathcal{C}}_1 + \mathcal{U}(1 + 1/\tau_\mathrm{c})$, $\mathcal{C}_2 = \mathcal{U}/\tau_\mathrm{c}$, and

$$\mathcal{U} = \frac{1}{\mu}\frac{P_0}{\Upsilon}\frac{\Gamma(\alpha/2+1)}{(\pi\lambda)^{\alpha/2}} \quad (13)$$

with $\mu$ accounting for the efficiency of the power amplifier (PA). The term $\mathcal{A}\mathsf{ASE}$ accounts for the power consumed per unit area for coding, decoding and backhaul, which is proportional to the ASE with proportionality constant $\mathcal{A} = P_\mathrm{DEC} + P_\mathrm{COD} + P_\mathrm{BT}$. The term $\mathcal{C}_0 = P_\mathrm{FIX} + P_\mathrm{SYN}$ models the static power consumption at each BS, and $\mathcal{C}_1 K$ with $\bar{\mathcal{C}}_1 = P_\mathrm{UE} + 5B_\mathrm{w}/(\tau_\mathrm{c}L_\mathrm{BS})$ accounts for the power consumed at the UEs as well as for part of the power consumed for ZF combining at the BS, which scales with $K$. The cubic term $\mathcal{C}_3 K^3$ shows up due to ZF with $\mathcal{C}_3 = B_\mathrm{w}/(\tau_\mathrm{c}L_\mathrm{BS})$. The term $\mathcal{D}_0 M$ with $\mathcal{D}_0 = P_\mathrm{BS}$ models the power consumed by the BS transceiver chain. Finally, $\mathcal{D}_1 MK$ and $\mathcal{D}_2 MK^2$ with $\mathcal{D}_1 = \frac{3B_\mathrm{w}}{\tau_\mathrm{c}L_\mathrm{BS}}\left(\frac{5}{2} + \tau_\mathrm{c}\right)$ and $\mathcal{D}_2 = 9B_\mathrm{w}/(2\tau_\mathrm{c}L_\mathrm{BS})$ account for part of the power consumed by signal processing for channel estimation and ZF combining.

TABLE I
NETWORK PARAMETERS

| Parameter | Value |
|---|---|
| Fixed power: $P_\mathrm{FIX}$ | 10 W |
| Power for BS Local Oscillator: $P_\mathrm{LO}$ | 0.2 W |
| Power per BS antennas: $P_\mathrm{BS}$ | 0.4 W |
| Power for antenna at UE: $P_\mathrm{UE}$ | 0.2 W |
| Power for data coding: $P_\mathrm{COD}$ | 0.1 W/(Gbit/s) |
| Power for data decoding: $P_\mathrm{DEC}$ | 0.8 W/(Gbit/s) |
| BS computational efficiency: $L_\mathrm{BS}$ | 75 Gflops/W |
| UE computational efficiency: $L_\mathrm{UE}$ | 3 Gflops/W |
| Power for backhaul traffic: $P_\mathrm{BT}$ | 0.25 W/(Gbit/s) |
| Path loss exponent: $\alpha$ | 3.76 |
| Coherence block length: $\tau_\mathrm{c}$ | 400 samples |
| Propagation loss at 1 km: $\Upsilon$ | 130 dB |
| Power amplifier efficiency: $\mu$ | 0.39 |

## IV. ENERGY EFFICIENCY MAXIMIZATION

Next, we analyze the EE maximization problem in (5). Notice that, due to the unavoidable inter-cell interference in cellular networks, (5) is only feasible if $\gamma < \tau_\mathrm{c}(\alpha - 1)$ as it follows from observing that (8) is monotonically increasing in $M$ and the largest value $\zeta(\alpha-1)$ is obtained when $M \to \infty$. If then $\zeta$ is set equal to its maximum value $\tau_\mathrm{c}$ (obtained for $K = 1$), the above condition is obtained. Notice that the problem is feasible in most cases of practical interest [9].

### A. Optimal Pilot Reuse Factor

We begin by computing the optimal pilot reuse factor when $M$ and $K$ are fixed.

**Lemma 2.** *Consider any pair of $(M,K)$ for which the problem (5) is feasible. The SINR constraint in (5) is satisfied by selecting*

$$\zeta^\star \geq \frac{B_1 \gamma}{M - K - B_2 \gamma} \quad (14)$$

*where*

$$B_1 = \frac{4K}{(\alpha-2)^2} + \frac{M}{\alpha-1} + \frac{2\left(K + \frac{1}{\mathsf{SNR}}\right)}{\alpha-2} - \frac{K}{\alpha-1} \quad (15)$$

$$B_2 = \left(\frac{1}{\mathsf{SNR}} + \frac{2K}{\alpha-2}\right)\left(1 + \frac{1}{\mathsf{SNR}_p}\right) + \frac{K}{\mathsf{SNR}_p}. \quad (16)$$

*Proof.* By gathering the terms that contain $\zeta$ and using (8), we obtain (14) by solving the resulting inequality for $\zeta$. □

The above lemma provides insights into how the EE-optimal pilot reuse factor $\zeta^\star$ depends on the other system parameters. In particular, it shows that $\zeta^\star$ must increase with $K$ to guarantee a certain average SINR. This is intuitive since increasing $\zeta$ leads to better channel estimation which can partially suppress the increased interference from having more UEs. Comparing (14) with the optimal pilot reuse factor

$$\zeta_{\text{MR}}^\star \geq \frac{B_1\gamma + \frac{2K}{\alpha-1}\gamma}{M - K\gamma - B_2\gamma} \quad (17)$$

obtained in [9] with MR combining, it follows that with MR the denominator scales with $K\gamma$ rather than $K$ and we have a positive extra term at the numerator. Since $\gamma$ is commonly greater than 1 (to ensure reasonable average SE constraints), it turns out that a smaller pilot reuse factor can be used with ZF due to its interference suppression capabilities. Notice that $\zeta^\star$ is a decreasing function of $M$, SNR and $\text{SNR}_p$. This is because all these parameters amplify the desired signal, which, as a consequence, improves the channel estimation and makes the system operate in a less noise limited regime. The pilot reuse factor $\zeta^\star$ reduces as the path loss exponent $\alpha$ increases (since $B_1$ and $B_2$ are reduced), which is natural since inter-cell interference decays more quickly. Next, to minimize the pilot overhead we assume that

$$\zeta^\star = \frac{B_1\gamma}{M - K - B_2\gamma}. \quad (18)$$

### B. Optimal Number of Antennas and UEs per BS

Plugging (18) into (5), the EE maximization problem reduces to

$$\underset{M,K\in\mathbb{Z}_{++}}{\text{maximize}} \quad \text{EE}(\zeta^\star) = \frac{B_{\text{w}}\,\text{ASE}(\zeta^\star)}{\mathcal{A}\,B_{\text{w}}\,\text{ASE}(\zeta^\star) + \text{APC}(\zeta^\star)} \quad (19)$$
$$\text{subject to} \quad 1 \leq \zeta^\star \leq \tau_c/K$$

with

$$\text{ASE}(\zeta^\star) = K\left(1 - \frac{K\zeta^\star}{\tau_c}\right)\log_2(1+\gamma) \quad (20)$$
$$\text{APC}(\zeta^\star) = \mathcal{C}_0 + \mathcal{C}_1 K - \mathcal{C}_2 K^2 \zeta^\star + \mathcal{C}_3 K^3$$
$$+ \mathcal{D}_0 M + \mathcal{D}_1 MK + \mathcal{D}_2 MK^2 \quad (21)$$

and $\zeta^\star$ given by (18). To find the optimal values for $M$ and $K$, an integer-relaxed version of (19) is first considered where $M$ and $K$ can be any positive scalars. The integer-valued solutions are then extracted from the relaxed problem. For analytic tractability, we replace $M$ with $\bar{c} = M/K$, which is the number of BS antennas per UE. This yields:

$$\underset{\bar{c},K\in\mathbb{Z}_{++}}{\text{maximize}} \quad \text{EE}(\zeta^\star) = \frac{B_{\text{w}}\,\text{ASE}(\zeta^\star)}{\mathcal{A}\,B_{\text{w}}\,\text{ASE}(\zeta^\star) + \text{APC}(\zeta^\star)} \quad (22)$$
$$\text{subject to} \quad 1 \leq \zeta^\star \leq \tau_c/K$$

with

$$\zeta^\star = \frac{B_1\gamma/K}{\bar{c} - 1 - B_2\gamma/K}. \quad (23)$$

For simplicity, we define the following quantities:

$$a_0 = \frac{K^2}{\alpha-1}\frac{\gamma}{\tau_c} \quad (24)$$
$$a_1 = \frac{\gamma K^2}{\tau_c}\left(\frac{4}{(\alpha-2)^2} + \frac{2}{\alpha-2} - \frac{1}{\alpha-1}\right) + \frac{2\gamma K}{\tau_c\text{SNR}(\alpha-2)} \quad (25)$$
$$a_2 = K \quad (26)$$
$$a_3 = K + \frac{\gamma K}{\text{SNR}_p} + \left(\frac{2\gamma K}{\alpha-2} + \frac{\gamma}{\text{SNR}}\right)\left(1 + \frac{1}{\text{SNR}_p}\right) \quad (27)$$
$$a_4 = \mathcal{D}_0 K + \mathcal{D}_1 K^2 + \mathcal{D}_2 K^3 \quad (28)$$
$$a_5 = \mathcal{C}_0 + \mathcal{C}_1 K + \mathcal{C}_3 K^3 \quad (29)$$
$$a_6 = \tau_c\mathcal{C}_2 K \quad (30)$$

and $r_0 = a_2 - a_0$, $r_1 = a_1 + a_3$, $q_0 = a_1 a_6 + a_3 a_5$, $q_1 = a_3 a_4 + a_0 a_6 - a_2 a_5$ and $q_2 = a_2 a_4$. For a given $K$, the EE-maximizing value of $\bar{c}$ is obtained as:

**Lemma 3.** *Consider the optimization problem (22) where $\bar{c} = M/K$ and $K$ are relaxed to be real-valued variables. For any fixed $K > 0$ such that the relaxed problem is feasible, the EE is maximized by*

$$\bar{c}^\star = \min\left(\max\left(\bar{c}', \bar{c}_1\right), \bar{c}_2\right) \quad (31)$$

*with*

$$\bar{c}' = \frac{r_1}{r_0} + \sqrt{-\frac{q_0}{q_2} - \frac{q_1}{q_2}\frac{r_1}{r_0} + \left(\frac{r_1}{r_0}\right)^2} \quad (32)$$
$$\bar{c}_1 = \frac{a_1 + a_3}{a_2 - a_0} \quad \text{and} \quad \bar{c}_2 = \frac{\frac{K}{\tau_c}a_1 + a_3}{a_2 - \frac{K}{\tau_c}a_0}. \quad (33)$$

*Proof.* We start computing the term $K\zeta^\star/\tau_c$ that appears both in the $\text{ASE}(\zeta^\star)$ and $\text{APC}(\zeta^\star)$. Plugging (15) and (16) into (23) yields $K\zeta^\star/\tau_c = (a_0\bar{c}+a_1)/(a_2\bar{c}-a_3)$ such that the constraint of problem (22) can be rewritten as

$$1 \leq \frac{a_0\bar{c} + a_1}{a_2\bar{c} - a_3} \leq \frac{\tau_c}{K} \quad (34)$$

or, equivalently, as $\bar{c}_1 \leq \bar{c} \leq \bar{c}_2$. The objective function of (22) reduces to

$$\frac{\left(1 - \frac{a_0\bar{c}+a_1}{a_2\bar{c}-a_3}\right)}{a_4\bar{c} + a_5 - a_6\frac{a_0\bar{c}+a_1}{a_2\bar{c}-a_3}} \quad (35)$$

which can be easily shown to be a quasi-concave function of $\bar{c}$. By taking the first derivative of (35) and equating to zero we obtain $\bar{c}'$ in (32), which corresponds to the solution of the unconstrained problem. This yields the desired result. □

This lemma shows how the optimal $\bar{c}$ depends on the other system parameters. In particular, we see that $\bar{c}^\star$ increases roughly linearly with $K$ and $\gamma$. This is reasonable since the network tends to equip the BSs with more antennas in order to guarantee an increase of the minimum average SINR to each UE. The same happens with respect to the circuit power parameters given by $\mathcal{C}_0 = P_{\text{FIX}} + P_{\text{SYN}}$ and $\bar{\mathcal{C}}_1 K = P_{\text{UE}} + 5B_{\text{w}}/(\tau_c L_{\text{BS}})$. In contrast, $\bar{c}^\star$ is inversely proportional to the BS density as $\lambda^{\alpha/4}$ (since $\mathcal{U}$ in $a_5$ is reduced as $\lambda^{-\alpha/2}$); fewer antennas must be used if the BS density increases. The same happens with respect to $\mathcal{D}_0 = P_{\mathrm{BS}}$ since it becomes more costly to have additional antennas when $P_{\mathrm{BS}}$ increases. Finally, fewer antennas are needed when SNR and $\mathrm{SNR}_p$ are increased.

If $\bar{c}$ is given, then the optimal $K$ is obtained as follows:

**Corollary 1.** *Consider the optimization problem* (22) *where $\bar{c} = M/K$ and $K$ are relaxed to be real-valued variables. For any fixed $\bar{c} > 0$ such that the relaxed problem is feasible, the optimal number of UEs is the root to a quintic equation.*

This corollary can be easily proved by taking the derivative of (22) with respect to $K$. Notice that there exist no generic closed-form root expressions for a quintic equation but solutions can be found accurately using numerical algorithms. To gain insights into how $K^\star$ is affected by the different parameters, assume that the power consumption required for linear processing and channel estimation are both negligible, which implies $\mathcal{C}_3 = \mathcal{D}_1 = \mathcal{D}_2 \approx 0$. This case is relevant as all these terms essentially decrease with the computational efficiency $L_{\mathrm{BS}}$, which is expected to increase rapidly in the future. We assume also that $\mathrm{SNR} \gg \gamma$ and the BS density is sufficiently large such that $\mathcal{C}_2$ is negligible (since $\mathcal{U}$ reduces as $\lambda^{-\alpha/2}$). Then, the following result is of interest:

**Corollary 2.** *Consider the optimization problem* (22) *where $\bar{c} = M/K$ and $K$ are relaxed to be real-valued variables. For any fixed $\bar{c} > 0$ such that the relaxed problem is feasible, if we let $\lambda \to \infty$, $L_{\mathrm{BS}} \to \infty$ and $\mathrm{SNR} \gg \gamma$, the optimal number of UEs is*

$$K^\star \approx \frac{\mathcal{C}_0}{\mathcal{C}_1 + \mathcal{D}_0 \bar{c}} \left( \sqrt{1 + \frac{b_2 - b_1}{b_0} \frac{\mathcal{C}_1 + \mathcal{D}_0 \bar{c}}{\mathcal{C}_0}} - 1 \right) \quad (36)$$

*where* $b_1 = \frac{\gamma}{\tau_c} \frac{2}{\alpha-2} \frac{1}{\mathrm{SNR}}$, *while $b_0$ and $b_2$ are given by*

$$b_0 = \frac{\gamma}{\tau_c} \left( \frac{4}{(\alpha-2)^2} + \frac{2}{\alpha-2} - \frac{1}{\alpha-1} + \frac{\bar{c}}{\alpha-1} \right) \quad (37)$$

$$b_2 = \bar{c} - 1 - \frac{2\gamma}{\alpha-2} \left( 1 + \frac{1}{\mathrm{SNR}_p} \right) - \frac{\gamma}{\mathrm{SNR}_p}. \quad (38)$$

*Proof.* If $\bar{c}$ is given and we let $\lambda$ and $L_{\mathrm{BS}}$ goes to infinity, then $\mathcal{C}_2 = \mathcal{C}_3 = \mathcal{D}_1 = \mathcal{D}_2 \approx 0$, then (22) reduces to a quasi-concave function whose maximum is achieved for (36). $\square$

The above result coincides with that in [9] and shows that, under the above circumstances, $K^\star$ decreases with $\bar{c}^\star$ as $\sqrt{1/\bar{c}^\star}$ when $\bar{c}^\star$ increases. From (36), it is found that $K^\star$ increases with the static energy consumption $\mathcal{C}_0$, while it decreases with $\mathcal{C}_1$ and $\mathcal{D}_0$. The same behavior is observed for the optimal number of BS antennas. Therefore, we may conclude that more BS antennas and UEs per cell can be supported only if the increase in circuit power has a marginal effect on the consumed power. In addition, we note that $K^\star$ is a decreasing function of $\gamma$, since the interference increases as the SINR target of the UEs increases.

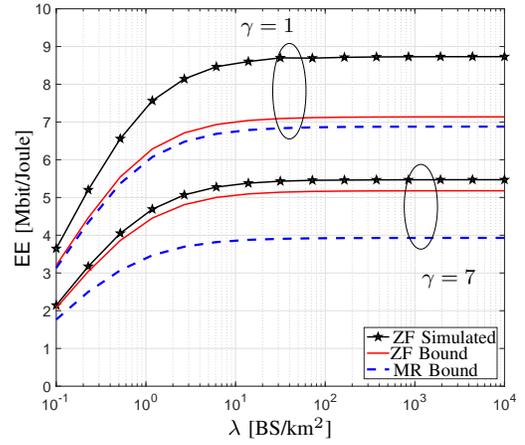

Fig. 1. Energy efficiency as a function of $\lambda$ for $\gamma = 1$ and 7. ZF is evaluated computing upper bounds from Monte-Carlo simulations (marked black line) and the lower bound (red line) and then compared to MR (dashed blue line).

To summarize, we first showed in Lemma 2 how to compute the optimal pilot reuse factor. Then, Lemma 3 and Corollary 2 showed how to optimize the EE separately with respect to $M$ and $K$. To solve the original problem (5) *jointly* with respect to all the parameters, we propose the following alternating optimization algorithm:

1) Assume that an initial feasible point $(\zeta, M, K)$ is given;
2) Optimize $\zeta$ by using Lemma 2;
3) Optimize $M$ by using Lemma 3;
4) Optimize $K$ by using Corollary 2;
5) Obtain the integer solutions from the real-valued ones;
6) Repeat 2) – 5) until convergence is achieved.

The convergence of the above algorithm is validated by numerical results in the sequel.

## V. NUMERICAL RESULTS

Monte Carlo simulations are now used to design the network for maximal EE and validate the analysis. The hardware as well as the channel parameters are taken from [12] and [9] and listed in Table I. We consider a 9 macro-cell setup scenario with wraparound topology. Each macro-cell is a square of area 1 km$^2$ wherein $\lambda$ BSs are deployed as described in Section II. The transmission bandwidth is $B_{\mathrm{w}} = 20$ MHz and each coherence block consists of $\tau_c = 400$ samples. We assume that $\mathrm{SNR}_p = 5$ dB and $\mathrm{SNR} = 0$ dB.

Fig. 1 shows the EE as a function of $\lambda$ for different SINR constraints $\gamma \in \{1, 7\}$, which corresponds to the average gross SEs $\log_2(1+\gamma) \in \{1, 3\}$. The pilot reuse $\zeta$ is chosen optimally according to (18). The EE is obtained as in (5) using the power model in (12) and the average ergodic SE either lower bounded as in Theorem 1, i.e., using $\underline{\mathrm{SE}}$, or computed by Monte Carlo simulations. Fig. 1 shows that the function is monotonically increasing with $\lambda$ making small-cells a promising solution for maximal-EE deployment. As it is seen, ZF outperforms MR for any value of $\lambda$. The gain reduces as $\gamma$ decreases since the two combining schemes tend to perform the same in a noise limited regime.

Fig. 2 shows the EE lower bound as a function of $K$ and $M > K$ for optimal $\zeta$ with $\gamma = 3$ and $\mathrm{SNR} = 0$ dB. As it is seen, the EE surface is pseudo-concave and has a unique global maximizer, which can be computed applying the alternating optimization algorithm in Section IV. The global maximizer is achieved by the triplet $(M^\star, K^\star, \zeta^\star) = (91, 10, 7.2)$ and gives a maximum of EE$^\star = 6.6$ Mbit/Joule. The ratio $M/K$ for the considered setup is approximately equal to 10, which leads to a massive MIMO setup. The pilot reuse probability $1/\zeta^\star = 13.9\%$ is low enough to ensure robustness against pilot contamination. Compared to [9], the results in Fig. 2 shows that ZF provides a smoother EE function than MR, which makes ZF more robust to system changes and thus a better choice as BS combining scheme. In Table II, ZF and MR are compared with $\gamma \in \{1, 7\}$ and $\lambda = 100$ BS/km$^2$. It turns out that, for a given $\gamma$, both schemes employ almost the same optimal number of antennas at each BS and serve approximately the same number of UEs. With both schemes, the ratio $M/K$ increases almost linearly with $\gamma$ as stated in Lemma 3. ZF achieves a higher EE than MR in the case that we study. This is a direct consequence of the ZF capabilities to handle intra-cell interference, which allows the BS to serve more UEs with a smaller number of antennas. This has a doubly-positive effect: higher ASE due to multiplexing gain and lower APC because of smaller arrays. Notice also that, as claimed in Corollary 2, $K^\star$ decreases with $\gamma$ since the interference increases as more UEs are served. The pilot reuse factor is smaller with MR. This happens because ZF mitigates the intra-cell interference and allows a higher inter-cell interference.

TABLE II
OPTIMAL NETWORK DESIGN PARAMETERS AND PERFORMANCE ACHIEVED FOR $\lambda = 100$ [BS/KM$^2$].

| Combiner | EE$^\star$ [Mbit/Joule] | ASE$^\star$ [kbit/s/Hz/km$^2$] | APC$^\star$ [kW/km$^2$] | $M^\star$ | $K^\star$ | $\zeta^\star$ | $1/\zeta^\star$ [%] |
|---|---|---|---|---|---|---|---|
| ZF ($\gamma = 1$) | 7 | 8.3 | 23.6 | 78 | 20 | 3.4 | 29.8 |
| ZF ($\gamma = 3$) | 6.6 | 8.2 | 24.9 | 91 | 10 | 7.3 | 13.8 |
| ZF ($\gamma = 7$) | 4.7 | 7.2 | 30.6 | 122 | 6 | 13.2 | 7.6 |
| MR ($\gamma = 1$) | 6.8 | 7.8 | 23 | 76 | 19 | 3.8 | 26 |
| MR ($\gamma = 3$) | 5.4 | 7.4 | 27.3 | 104 | 9 | 7.9 | 12.6 |
| MR ($\gamma = 7$) | 3.6 | 6.1 | 33.8 | 139 | 5 | 14.7 | 6.8 |

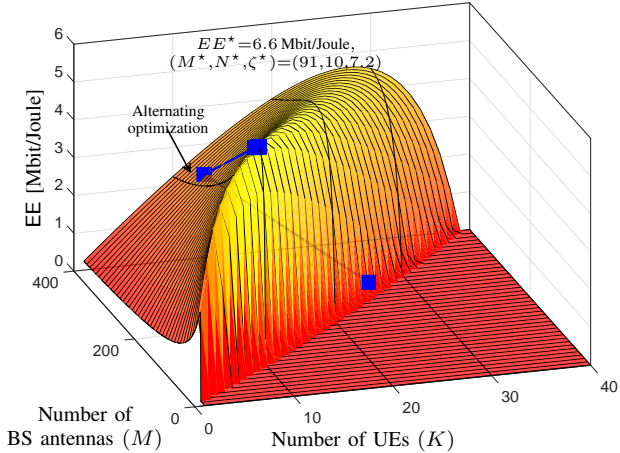

Fig. 2. Energy efficiency for different values of $M$ and $K$ for fixed $\gamma = 3$ and $\lambda = 100$ BS/km$^2$.

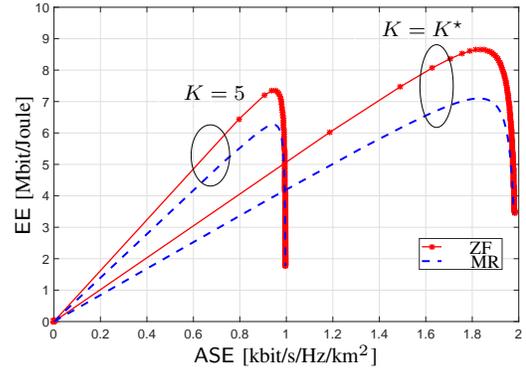

Fig. 3. Energy efficiency as a function of the area spectral efficiency for different $K = 5, 10$ when $\gamma = 3$.

Fig. 3 illustrates the EE versus the ASE of ZF and MR combining for $K = \{5, K^\star\}$ when $\lambda = 100$ BS/km$^2$ and $\gamma = 3$. The pilot reuse $\zeta$ is chosen optimally according to (18). The SE is computed using the lower bound in Theorem 1. As it is seen, the EE is a unimodal function of the ASE. This means that there exist operating conditions under which it is possible to jointly increase SE and EE. Also, ZF performs always better than MR, though it assures approximately the same ASE as MR at the EE-optimal point.

## VI. CONCLUSIONS

We have designed a cellular network for maximal energy efficiency in the UL when ZF is used for data detection. This was formulated as an optimization problem by using stochastic geometry, a new lower bound on the SE, and a state-of-the-art power consumption model. The variables were pilot reuse factor, number of BS antennas and UEs per BS. The results showed that ZF allows a higher densification of the network and the use of a smaller pilot reuse factor while achieving a higher EE than with MR combining. Also, it turned out that the EE-optimal configuration resembles a Massive MIMO setup.

## APPENDIX A

The average ergodic UL channel capacity of the typical UE $k$ in cell $j$ can be lower bounded by $(1 - K\zeta/\tau_c)\mathbb{E}_d\{\log_2(1 + \mathsf{SINR}_{jk})\}$ where $\mathsf{SINR}_{jk}$ is given in (39) on the top of the next page. The expectations $\mathbb{E}_{\{\mathbf{h}, a\}}\{\cdot\}$ are computed with respect to channel realizations and pilot allocations. The outer expectation $\mathbb{E}_d\{\cdot\}$ computes the average over UEs positions. The above lower bound is valid for any choice of the receive combining scheme. As shown in [9],

$$\text{SINR}_{jk} = \frac{p_{jk}|\mathbb{E}_{\{\mathbf{h},a\}}\{\mathbf{v}_{jk}^{\text{H}}\mathbf{h}_{jk}^{j}\}|^2}{\sum_{l\in\Phi_\lambda}\sum_{i=1}^{K} p_{li}\mathbb{E}_{\{\mathbf{h},a\}}\{|\mathbf{v}_{jk}^{\text{H}}\mathbf{h}_{li}^{j}|^2\} - p_{jk}|\mathbb{E}_{\{\mathbf{h},a\}}\{\mathbf{v}_{jk}^{\text{H}}\mathbf{h}_{jk}^{j}\}|^2 + \sigma^2\mathbb{E}_{\{\mathbf{h},a\}}\{\|\mathbf{v}_{jk}\|^2\}} \quad (39)$$

$$\mathbb{E}_d\{\underline{\text{SINR}}_{jk}^{-1}\} = \frac{1}{M-K}\left(\left(K+\frac{1}{\text{SNR}}\right)\left(1+\frac{1}{\text{SNR}_p}\right) + \left(K+\frac{1}{\text{SNR}}\right)\frac{1}{\zeta}\mathbb{E}_d\{\vartheta_{jk}^{(1)}\} + \left(1+\frac{1}{\text{SNR}_p}\right)\sum_{i=1}^{K}\mathbb{E}_d\{\vartheta_{ji}^{(1)}\}\right.$$
$$\left. + \frac{1}{\zeta}\sum_{i=1}^{K}\mathbb{E}_d\{\vartheta_{jk}^{(1)}\vartheta_{ji}^{(1)}\} + \frac{M-K}{\zeta}\mathbb{E}_d\{\vartheta_{jk}^{(2)}\} - \frac{1}{\zeta}\sum_{i=1}^{K}\mathbb{E}_d\{\vartheta_{ji}^{(2)}\} - K\right) \quad (40)$$

closed-form expressions for all terms in (39) can be computed for MR combining. Next, we show that the same can be done for ZF. Before doing this, we observe that

$$\mathbb{E}_{\{a\}}\left\{\frac{1}{\gamma_{jk}^j}\right\} = \frac{1}{\beta_{jk}^j}\left(1 + \frac{\vartheta_{jk}^{(1)}}{\zeta} + \frac{1}{\text{SNR}_p}\right) = \frac{1}{\beta_{jk}^j}A_{jk}^{(1)} \quad (41)$$

as it follows from (3) taking into account that $\mathbb{E}_{\{a\}}\{a_l^j\} = 1/\zeta$ by design. Also, we have that

$$\mathbb{E}_{\{a\}}\left\{\frac{\gamma_{ji}^j}{\gamma_{jk}^j}\right\} \stackrel{(a)}{\geq} \frac{\mathbb{E}_{\{a\}}\{\gamma_{ji}^j\}}{\mathbb{E}_{\{a\}}\{\gamma_{jk}^j\}} \stackrel{(b)}{=} \frac{\beta_{ji}^j}{\beta_{jk}^j} \quad (42)$$

where $(a)$ is direct consequence of Jensen's inequality and the independence of $\gamma_{jk}^j$ and $\gamma_{ji}^j$ since they cannot share the same pilot sequence, and $(b)$ follows from (3).

The useful term is computed as

$$|\mathbb{E}_{\{\mathbf{h},a\}}\{\mathbf{v}_{jk}^{\text{H}}\mathbf{h}_{jk}^{j}\}|^2 \stackrel{(a)}{=} |\mathbb{E}_{\{\mathbf{h},a\}}\{\mathbf{v}_{jk}^{\text{H}}\hat{\mathbf{h}}_{jk}^{j}\}|^2 \stackrel{(b)}{=} 1 \quad (43)$$

where $(a)$ follows from $\mathbf{h}_{jk}^j = \hat{\mathbf{h}}_{jk}^j + \tilde{\mathbf{h}}_{jk}^j$ and the fact that $\hat{\mathbf{h}}_{jk}^j$ and $\tilde{\mathbf{h}}_{jk}^j$ are independent and $(b)$ comes from ZF combining. The noise term is obtained as

$$\mathbb{E}_{\{\mathbf{h},a\}}\{\|\mathbf{v}_{jk}\|^2\} \stackrel{(a)}{=} \frac{1}{M-K}\mathbb{E}_{\{a\}}\left\{\frac{1}{\gamma_{jk}^j}\right\} \stackrel{(b)}{=} \frac{1}{\beta_{jk}^j}\frac{A_{jk}^{(1)}}{M-K} \quad (44)$$

where $(a)$ follows from the statistics of $\hat{\mathbf{h}}_{jk}^j$ and the properties of Wishart matrices (e.g., [7, Proof of Proposition 3]) and $(b)$ is obtained using (41).

Consider now the cell of interest $j$. Then, by using the inequality in (42), for any UE $i$ we have

$$\mathbb{E}_{\{\mathbf{h},a\}}\{|\mathbf{v}_{jk}^{\text{H}}\mathbf{h}_{ji}^{j}|^2\} = \delta(i-k) + \frac{\beta_{ji}^j}{\beta_{jk}^j}\frac{A_{jk}^{(1)}-1}{M-K}. \quad (45)$$

Consider all the UEs that do not cause pilot contamination to UE $k$ of cell $j$ (i.e., $a_l^j = 0$). Then, we have that $\mathbb{E}_{\{\mathbf{h}\}}\{|\mathbf{v}_{jk}^{\text{H}}\mathbf{h}_{li}^{j}|^2\} = \beta_{li}^j\mathbb{E}_{\{\mathbf{h},a\}}\{\|\mathbf{v}_{jk}\|^2\}$, which follows from the fact that $\mathbf{v}_{jk}$ is independent from $\mathbf{h}_{li}^j$ in presence of no pilot contamination. As for the UEs that cause pilot contamination (i.e., $a_l^j = 1$), we have that

$$\mathbb{E}_{\{\mathbf{h}\}}\{|\mathbf{v}_{jk}^{\text{H}}\hat{\mathbf{h}}_{li}^{j}|^2\} = \frac{(\beta_{li}^j)^2}{\beta_{li}^l\beta_{ji}^j}\delta(i-k)$$

$$\mathbb{E}_{\{\mathbf{h}\}}\{|\mathbf{v}_{jk}^{\text{H}}\tilde{\mathbf{h}}_{li}^{j}|^2\} = \beta_{li}^j\mathbb{E}_{\{\mathbf{h},a\}}\{\|\mathbf{v}_{jk}\|^2\} - \frac{1}{M-K}\frac{(\beta_{li}^j)^2}{\beta_{li}^l\beta_{ji}^j}.$$

By using the above results into (39) we obtain (4).

## APPENDIX B

Next, a tractable lower bound on the average SE given by $(1 - K\zeta/\tau_c)\mathbb{E}_d\{\log_2(1 + \underline{\text{SINR}}_{jk})\}$ is computed, where the expectation is taken with respect to the PPP and UE locations. By using the Jensen's inequality, we obtain $\mathbb{E}_d\{\log_2(1+\underline{\text{SINR}}_{jk})\} \geq \log_2(1 + 1/\mathbb{E}_d\{\underline{\text{SINR}}_{jk}^{-1}\})$. From (4), the expectation of $\underline{\text{SINR}}_{jk}^{-1}$ can be expanded as in (40). Using the results below, taken from [9],

$$\mathbb{E}_d\left\{\sum_{l\in\Psi_\lambda}\left(\frac{d_{li}^l}{d_{li}^j}\right)^{\kappa\alpha}\right\} = \frac{2}{\kappa\alpha - 2} \quad (46)$$

$$\mathbb{E}_d\left\{\sum_{n\in\Psi_\lambda}\sum_{l\in\Psi_\lambda\setminus\{n\}}\left(\frac{d_{nk}^n}{d_{nk}^j}\right)^\alpha\left(\frac{d_{li}^l}{d_{li}^j}\right)^\alpha\right\} = \frac{4}{(\alpha-2)^2} \quad (47)$$

$$\mathbb{E}_d\left\{\sum_{n\in\Psi_\lambda}\left(\frac{d_{nk}^n}{d_{nk}^j}\right)^\alpha\left(\frac{d_{ni}^n}{d_{ni}^j}\right)^\alpha\right\} \leq \frac{1}{\alpha-1} \quad (48)$$

we obtain (8). The details are skipped for space limitations.